\documentclass[12pt]{iopart}
\usepackage{graphics,iopams}

\newcommand{\dasgraph}[1]{\centerline{\includegraphics{#1}}}
\newcommand{\unit}[1]{\,\mbox{#1}}

\newcommand{\TiSa}{Ti:Sa}
\newcommand{\Rb}{$^{85}$Rb}
\newcommand{\ten}[2]{#1\times 10^{#2}}
\begin{document}
\title{Resolved diffraction patterns from a reflection grating for atoms}
\author{J Est\`eve, D Stevens, V Savalli, N Westbrook, C I Westbrook and A Aspect}
\address{Laboratoire Charles Fabry de l'Institut d'Optique, UMR 8501 du CNRS, 91403 Orsay Cedex, France}
\ead{jerome.esteve@iota.u-psud.fr}

\begin{abstract}
We have studied atomic diffraction at normal incidence from an
evanescent standing wave with a high resolution using velocity
selective Raman transitions. We have observed up to 3 resolved
orders of diffraction, which are well accounted for by a scalar
diffraction theory. In our experiment the transverse coherence
length of the source is greater than the period of the diffraction
grating.
\end{abstract}

\submitto{\JOB}

\maketitle
Diffraction of atoms by periodic potentials has been a subject of
study ever since the beginning of the field of atom optics. The
most extensive work has been done in a transmission
geometry~\cite{gould:1986,martin:1987,martin:1988,gould:1991,chapman:1995,
giltner:1995b,kunze:1996,kunze:1997,bernet:1996,billbragg,kokorowski:2001},
but reflection gratings have been demonstrated as well. Using
evanescent wave mirrors, reflection gratings have been produced by
retroreflecting the laser beam creating the evanescent
wave~\cite{brouri:1996,landragin:1997,cognet:1998,christ:1994} or
by temporally modulating the intensity of the evanescent
wave~\cite{steane:1996}. Magnetic mirrors have also been rendered
periodic by adding an appropriate external magnetic
field~\cite{rosenbusch:2000}.

Many early experiments on atomic reflection gratings were done at
grazing incidence, but this problem proved considerably more
subtle than was first imagined (for a review
see~\cite{henkel:1999}). It turned out that atomic diffraction at
grazing incidence cannot be analogous to the diffraction of light
(or X rays) on a reflection grating, because the reflection is not
on a hard wall, but on a soft barrier, so that the atom averages
out the grating modulation during the bounce. Alternatively, one
can  show that scalar ({\it i.e.} without internal state change)
atomic diffraction at grazing incidence on a soft reflection
grating would not allow energy and momentum conservation. This is
why grazing incidence atomic diffraction must be a process with an
internal state change. The change in internal energy allows the
process to conserve energy. ``Straightforward'', scalar
diffraction which is due only to a periodic modulation of the
atomic wave front, can only be achieved at normal incidence, as
implemented in reference~\cite{landragin:1997}.

This experiment suffered however from a resolution which was not
quite high enough to separate the individual diffraction peaks.
Although it was possible to deconvolve the experimental
resolution, and make a quantitative comparison with a scalar
diffraction theory, it was disappointing not to be able to
directly see the diffraction peaks. Here we discuss an improved
version of this experiment in which we use velocity selective
Raman transitions to select and analyze a velocity class much
narrower than what was possible in the experiments of
reference~\cite{landragin:1997}. We are thus able to resolve the
diffraction peaks.

Our results are also significant from the point of view of
measurements of the roughness of atomic mirrors. These
measurements have been performed for both magnetic and evanescent
wave
mirrors~\cite{landragin:1996a,voigt:2000,saba:1999,cognet:1999,savalli:2001}.
Some of the interpretation of these measurements relies on
theoretical treatments developed in close analogy with the theory
of atom diffraction~\cite{henkel:1997}, and it is clearly useful
to make additional tests of the theory when possible.

\begin{figure}[htbp]
\dasgraph{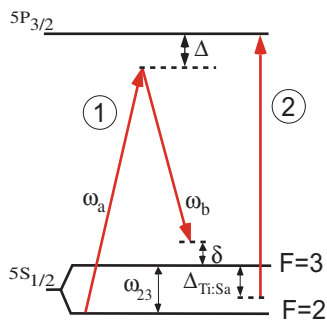} \caption{Level diagram for velocity
selective Raman transitions and evanescent wave atomic mirror. 1)
Frequencies $\omega_a$ and $\omega_b$ of the two
counterpropagating laser beams are separated by \Rb\ hyperfine
frequency $\omega_{23}$ plus Raman detuning $\delta$. 2) The
\TiSa\ laser that creates the evanescent wave mirror is blue
detuned for atoms in $F=3$ and red detuned for atoms in $F=2$,
thus only atoms in $F=3$ bounce on the atomic mirror. }
\label{fig.lasers}
\end{figure}

Our velocity selection scheme is shown in figure~\ref{fig.lasers}.
Two counterpropagating, phase-locked lasers induce transitions
between the two hyperfine levels of \Rb. They are detuned from
resonance by about 600~MHz ($\Delta$ in figure~\ref{fig.lasers}).
The detuning $\delta$ for this two photon transition depends on
the Doppler shift $\delta = k_{\mathrm{L}} (v \pm
v_{\rm{R}})$~\cite{kasevich:1991}, where $k_{\mathrm{L}}$ is the
magnitude of the laser wavevector and $v_{\rm{R}}$ the photon
recoil velocity. Thus only atoms with a given velocity $v$ such
that $\delta=0$ are resonant. The width of the velocity selection
is proportional to $1/\tau$, where $\tau$ is the duration of the
pulse. This width can be much smaller than the photon recoil
velocity. Figure \ref{fig.lasers} shows the various laser
frequencies used for the Raman transitions and for the evanescent
wave atomic mirror.

Our experiment follows closely the procedures of
reference~\cite{savalli:2001}. It consists of two distinct
sequences, each using two Raman pulses. First, to measure the
atomic velocity distribution after reflection, the following
sequence is used. A sample of \Rb\ atoms are optically pumped into
F=2 after preparation in a magneto-optical trap and optical
molasses, and released at time $t=0$. A constant magnetic field of
750~mG is turned on, to define the axis of quantization and split
the $m_F$ levels. A Raman $\pi$-pulse is applied at $t=8$~ms with
detuning $\delta_S$ (selection). This transfers atoms from
$|2,0,\delta_S / k_{\mathrm{L}} - v_{\rm{R}}\rangle$ into
$|3,0,\delta_S / k_{\mathrm{L}} + v_{\rm{R}} \rangle$\footnote{The
basis is $|F,m_F,v \rangle$ where $v$ is the atom velocity along
the laser beam propagation. }\footnote{We choose the hyperfine
sublevel $m_F=0$ because it is insensitive (to first order) to
magnetic fields.}. At $t=47$~ms, the \TiSa\ laser that forms the
evanescent wave is turned on. Its detuning is chosen so that it
reflects atoms in F=3 (blue detuning), while being attractive for
the state F=2 (red detuning). This eliminates unselected atoms
left in F=2. The \TiSa\ laser is turned off at $t=67$~ms. At
$t=120$~ms, a second $\pi$-pulse is applied, with detuning
$\delta_A$ (analysis), returning atoms in $|3,0,\delta_A /
k_{\mathrm{L}} + v_{\rm{R}}\rangle$ to $|2,0,\delta_A /
k_{\mathrm{L}} - v_{\rm{R}} \rangle$. At $t=124$~ms we switch on a
pushing beam for 2~ms, resonant for F=3, which accelerates F=3
atoms away from the detection region. Finally, the measured signal
is the fluorescence of the remaining atoms in F=2, illuminated by
a resonant probe beam with repumper and measured with a
photomultiplier tube. To measure the number of atoms at the end of
the sequence, we evaluate the area of a Gaussian fit of the
fluorescence signal.

This first sequence is repeated fixing $\delta_S$ and scanning
$\delta_A$. In this way, we build up the velocity distribution of the
reflected atoms. The frequencies $\delta_A$ are selected in a
random order, and after each shot, a second shot is taken with
$\delta_S = \delta_B$ (background) sufficiently detuned so that it
does not transfer any atoms.
Since we detect all atoms in F=2, there is a small background of
atoms that have not been selected, but instead
scattered a photon, thereby finishing in the same state as the
atoms that have been selected. This can occur during the analysis
pulse, and during interaction with the \TiSa\ laser. We estimate
that around $10\%$ of atoms undergo such processes in the
evanescent wave, and $1\%$ during each $\pi$ pulse. Most of these
atoms are accounted for by the background shot.
The background shot is subtracted from the signal shot,
and the difference is the data point.
We average each data point 5 times, so that a diffraction pattern takes around 20
minutes to acquire.

To complete the experiment, we need to know the velocity
distribution of the atoms selected by the first pulse. To measure
this we employ a second sequence, essentially the same as the
first, but without the mirror. Atoms are prepared in $F=3$ at the
end of the optical molasses. At $t=8\unit{ms}$ the first Raman
pulse transfers resonant atoms to $F=2$, other atoms are removed
with the pushing beam. At $t=22\unit{ms}$ the second Raman pulse
transfers atoms back in $F=3$ and we finally detect only atoms in
$F=3$. By scanning the detuning of the second Raman pulse we can
measure the velocity width of the atoms selected by the first
pulse. The observed velocity distribution is well fitted by a
Gaussian function with a width given by $2 \, k_{\mathrm{L}} \,
\sigma_{v} \approx 2 \pi/\tau$. Because of the convolution of the
selection and analysis pulses, the observed velocity distribution
is approximately $\sqrt 2$ times larger than that due to the
selection pulse alone. For a pulse length of $\tau=120$~$\mu$s,
the measured rms width is $2 \, k_{\mathrm{L}} \, \sigma_v = 2\pi
\times 7.7$~kHz or $\sigma_v = 0.50 \, v_{\rm{R}} =
3.0$\unit{mm.s$^{-1}$}. We will use this Gaussian function as our
resolution function $S(\delta)$. Any differences between the two
measured velocity distributions are attributed to the action of
the diffraction grating.



The reflection grating for atoms is made by retroreflecting a small
proportion of light that makes the evanescent wave: a standing
wave is created on the mirror and the potential seen by the atoms
can be written (neglecting the van der Waals interaction for the
moment)
\begin{equation}
U(x,z)=U_0 e^{-2 \, \kappa \, z}(1+ \epsilon \cos (2 \, k_{x} \,
x)),
\end{equation}
where $\epsilon$ is the contrast of the interference pattern,
$1/\kappa$ is the decay length of the electric field, and $k_{x}$ is
the propagation wave vector of the evanescent wave.
The contrast $\epsilon$ can be
written:
\begin{equation}
\epsilon = \frac{2 \sqrt{R}}{1+R}
\frac{k_{x}^2-\kappa^2}{k_{x}^2+\kappa^2},
\end{equation}
where $ R=I_{R}/I_{0}$ is the intensity fraction of the laser beam
that is retroreflected. To a good approximation, the grating
behaves as a thin phase grating, producing a modulated de Broglie
wave whose predicted phase modulation index is:
\begin{equation}
\varphi = \frac{2 \pi}{ \lambda_{\rm{dB}}}\frac{\epsilon}{\kappa},
\end{equation}
where $\lambda_{\rm{dB}}$ is the de Broglie wavelength of the
atomic cloud incident on the mirror. The van der Waals force
introduces a multiplicative correction that here increases the
phase modulation index $\varphi$~\cite{landragin:1997} by 12\% .

In our experiments, $\kappa=1.14 \, k_{\rm{L}}$. When the contrast
of the grating $\epsilon$ is of order $\lambda_{\rm{dB}}/\lambda_{\rm{L}}$,
we have $\varphi \sim 1$~rad, and the diffraction is efficient. Here,
$\lambda_{\rm{dB}} = 7$~nm and $\lambda_{\rm{L}} = 780$~nm. Thus
we only need a very weak contrast $\epsilon \sim 10^{-2}$,
corresponding to $R \sim 10^{-4}$.

The $n$th diffraction order acquires a transverse momentum of $2
\, n \, \hbar \, k_x $ and has an intensity of $J_n^2(\varphi)$,
where $J_n$ is the Bessel function of order $n$. Here, $k_x = 1.5
\, k_{\rm{L}}$, so to distinguish two successive diffraction
orders, we must have a velocity selection in the $x$ direction
better than $1.5 \, v_{\rm{R}}$.


\begin{figure}
\dasgraph{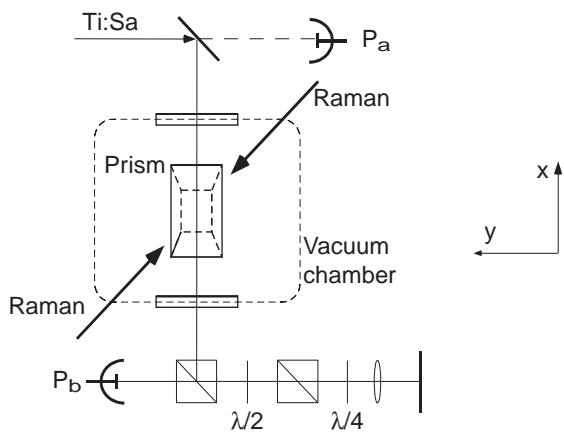} \caption{Experimental geometry. The \TiSa\
strikes the prism from below, with an angle of incidence
$\theta=53^\circ$.  The system of polarizers and wave plates
allows us to reflect a small controlled amount of the \TiSa\
beam.} \label{fig.setup}
\end{figure}

Figure \ref{fig.setup} shows the experimental geometry. The mirror
is 20.7~mm below the magneto-optical trap. It is a prism with an
index of refraction of $n=1.869$ and the angle of incidence of the
\TiSa\ laser $\theta = 53^{\circ}$. The prism is superpolished and
has a roughness of about $0.07$~nm deduced from measurements made
with an atomic force microscope and a Zygo optical heterodyne
profiler. To remove the degeneracy of the Zeeman sublevels, a
magnetic bias field of 750~mG is applied along the direction of
propagation of the Raman beams. After the first $\pi$ pulse, the
direction of the magnetic bias field is rotated by $45^{\circ}$ so
that it lies along the $y$-direction, corresponding to the
quantization axis of the evanescent wave. The change is carried
out over 10~ms, so that the atoms adiabatically follow the field.
The direction of the field is turned back after the bounce for the
second $\pi$ pulse.

Also shown in figure \ref{fig.setup} is the setup for measuring
and controlling the retroreflected power $R$. The two wave plates
control the power, the half wave plate is used as a coarse and the
quarter wave plate as a fine adjustment. A cat's eye arrangement
is used to control the spatial superposition and size of the
return beam. The experimental setup for the Raman beams has been
described elsewhere~\cite{savalli:2001}.

To measure $R$, first the incident power is found from $P_a$ and
the (known) transmission $T$ of the first mirror in figure
\ref{fig.setup}. Then the quarter wave plate is turned to the
position that maximizes the power $P_b$. Next, the position of the
half wave plate is fixed so that $P_b / (P_a / T)$ is the maximum
value of $R$ desired. Finally, the quarter wave plate is adjusted
to vary the $R$ over the desired range. The fraction of power
reflected $R$ is then deduced from the relationship
\begin{equation}
    R = \frac{P_b^{\mathrm{max}}}{P_a/T}\frac{\left( P_b^{\mathrm{max}} -
    P_b\right)}{P_b^{\mathrm{max}}}.
\end{equation}


\begin{figure}
\dasgraph  {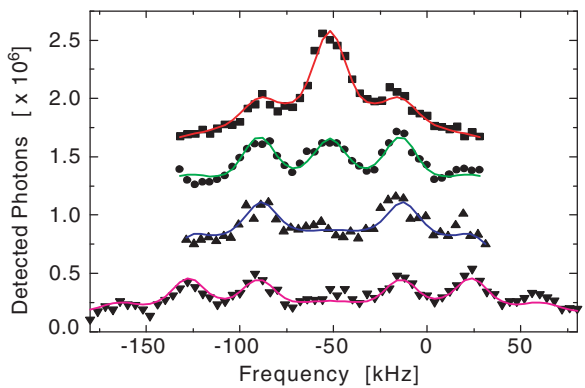} \caption{Velocity profile of diffracted
atoms (curves have been offset for display purposes). The
amplitude of the diffraction grating is increased from top to
bottom ($R = \ten{2.8}{-5}$, $\ten{5.7}{-5}$, $\ten{1.7}{-4}$,
$\ten{1.5}{-3}$). We can observe up to 3 separated orders of
diffraction.} \label{fig.diffract4}
\end{figure}

Figure \ref{fig.diffract4} (a,b,c,d) presents the atomic
experimental velocity distribution after reflection when $R =
\ten{2.8}{-5}$, $\ten{5.7}{-5}$, $\ten{1.7}{-4}$, $\ten{1.5}{-3}$
of the incident laser intensity is retroreflected, for
$\Delta_{Ti:Sa}=2.5$~GHz. The diffraction orders are more and more
populated when the intensity of the retroreflected beam increases,
and the different diffraction orders are well resolved. The
expected separation between successive orders of diffraction is
$4(\cos \alpha) \, k_x \, v_{\rm{R}} = 2 \pi \times 33$~kHz, where
$\alpha = 43^\circ$ is the measured angle between the Raman beams
and the evanescent wave propagation direction. The data in the
figure give a peak separation of $\delta_0 = 2 \pi \times 38$~kHz.
The difference between the calculated and observed peak
separations is rather large, but we currently have no satisfactory
explanation for this discrepancy.

The data are fitted by the following function:
\begin{equation}
 a + b \, \mathrm{R}(\delta) \times
     \sum_n \left[J_n^2(\phi) \, \delta(\delta-n \, \delta_0) \ast
           \left\{\mathrm{S}(\delta) +
            c \, \mathrm{M}(\delta)\right\} \right]
\end{equation}
Each diffraction order has a weight given by the value of the
Bessel function $J_n^2(\phi)$ and is convolved by the resolution
function $\mathrm{S}(\delta)$, plus the function
$\mathrm{M}(\delta)$ that accounts for a background of atoms whose
velocity distribution is imposed by the spatial extent of the
mirror. The resolution function is a gaussian profile with rms
width 7.7~kHz. This entire function is then multiplied by the
function $\mathrm{R}(\delta)$ which takes account of the spatial
variation of the Raman beams.

The widths of the functions $\mathrm{R}$, $\mathrm{S}$ and
$\mathrm{M}$ are calculated from measured values, so that there
are only four adjustable parameters for the fitting procedure: an
offset $a$, the amplitude of the curve $b$, the phase modulation
index $\phi$, and the background term $c$.
\begin{figure}
\dasgraph  {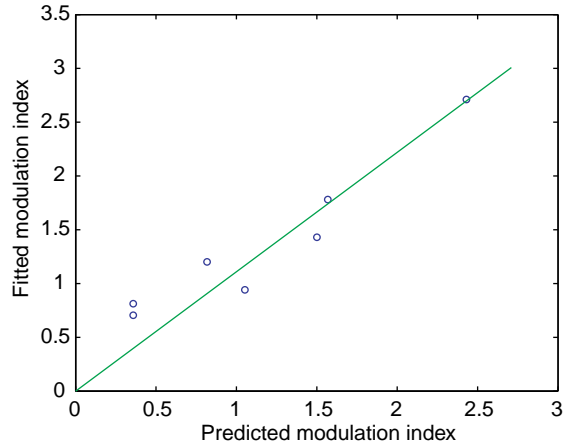} \caption{Fitted and predicted phase
modulation index. The fitted values are deduced from the velocity
profile of the diffracted atoms and the predicted ones are
calculated from the measurement of the amount of reflected power
$R$ (taking into account the Van der Waals interaction). The
straight line fit has gradient 1.11(8).} \label{fig.phi}
\end{figure}
Figure \ref{fig.phi} shows the fitted and predicted phase
modulation indices. The two are in satisfactory agreement to
within our uncertainty, which is chiefly limited by the
uncertainty in the superposition of the return beam.




To conclude, the fact that the peak heights can be accounted for
by a series of Bessel functions depending on a single parameter,
the modulation index, is quite remarkable and constitutes the
central result of this work. It shows that one can reach a regime
in which the atomic diffraction on a corrugated mirror can be
described simply as a reflection of a scalar matter wave, with a
phase modulation of the reflected wavefront calculated within the
straightforward thin phase grating
approximation~\cite{henkel:1994}. In this regime, one thus have a
close analogy to light diffraction off a reflection grating.

This work was supported by the European Union under grants
IST--1999--11055 and HPRN--CT--2000--00125, and by the DGA grant
00.34.025.

\ \

\end{document}